\documentclass[12pt]{article}
\usepackage{amsmath, hyperref, wasysym}
\usepackage{latexsym}
\usepackage{amssymb}
\usepackage{amsthm}
\usepackage[margin=.75in]{geometry}
\usepackage{wasysym}
\usepackage{hyperref}
\usepackage{times}
\usepackage{glossaries}
\usepackage{fancyhdr}
\usepackage{enumitem}
\usepackage{wrapfig}
\usepackage{times}
\usepackage{placeins}
\usepackage{graphicx}
\usepackage{cite}
\usepackage{pifont}
\usepackage{multicol}
\usepackage{color}
\usepackage{float}
\usepackage{wrapfig}
\usepackage{lipsum}
\restylefloat{table}
\usepackage{lineno}
\usepackage{authblk}
\usepackage[dvipsnames]{xcolor}
\usepackage{booktabs}
\usepackage[labelfont=bf]{caption}
\renewcommand{\figurename}{Figure}

\usepackage[utf8]{inputenc}

\makeglossaries
\newglossaryentry{pigment cell}
{
    name={\color{RubineRed}pigment cell},
    description={Cells are small units that all living things are made of (we are made of cells, and so is fish skin). Pigment cells are special cells that contain pigment (color)}
}

\newglossaryentry{model}
{
    name={\color{RubineRed}mathematical model},
    description={Rules or equations that describe something in a mathematical way. In this paper, the rule ``One side different, the rest the same" is part of our mathematical model}
}

\newglossaryentry{simulate}
{
    name={\color{RubineRed}simulate},
    description={To imitate a real-world system on a computer. You can simulate fish patterns on this \href{https://simulatingzebrafish.gitlab.io/zebrafish-outreach/}{\color{RubineRed}website}\footnote{Simulate a fish pattern here: \url{https://simulatingzebrafish.gitlab.io/zebrafish-outreach/}} with computer coding and a mathematical model, which provides directions to the computer}
}

\newglossaryentry{genes}
{
    name={\color{RubineRed}genes},
    description={The biological instructions that an animal gets from their parents to specify their features (like stripe or spot patterns for a fish, or green or brown eyes for a person)}
}

\newglossaryentry{mutation}
{
    name={\color{RubineRed}mutation},
    description={A change in an animal's genes from the normal setting to something else. Zebrafish usually have stripes (Figure~\ref{fig:zebrafishIntro}g), but mutant zebrafish (Figures~\ref{fig:zebrafishIntro}c, \ref{fig:zebrafishIntro}d) have maze or spot patterns}
}

\title{\color{ForestGreen}What's math got to do with patterns in fish?}
\author[1]{Blake Shirman}
\author[2,*]{Alexandria Volkening}
\affil[1]{Department of Mathematical Sciences, DePaul University, Chicago, IL, USA}
\affil[2]{Department of Mathematics, Purdue University, West Lafayette, IN, USA}
\affil[*]{Correspondence: \href{mailto:avolkening@purdue.edu}{avolkening@purdue.edu}}
\date{\today}

\usepackage{graphicx}

\begin{document}

\maketitle

\begin{abstract}
    When you think of fish, what comes to mind? Maybe you think of pet goldfish, movie characters like Dory or Nemo, or trout in a local river. One of the things that all of these fish have in common is patterns in their skin. Nemo sports black and white stripes in his orange skin, and trout have spots. Even goldfish have a pattern --- it's just plain gold (and kinda boring). Why do some fish have stripes, others have spots, and others have plain patterns? It turns out that this is a tricky question, so scientists need tools from several subjects to answer it. In this paper, we use biology, math, and computer coding to help figure out how fish get different skin patterns.
\end{abstract}

\noindent \paragraph{\textbf{Keywords}}{ ~math $|$ model $|$ zebrafish $|$ biology $|$ computer science $|$ coding $|$ pattern $|$ animal skin}

\section*{\color{ForestGreen}What are fish patterns made of?}

\begin{figure}[h!]
\centering
\includegraphics[width=0.8\textwidth]{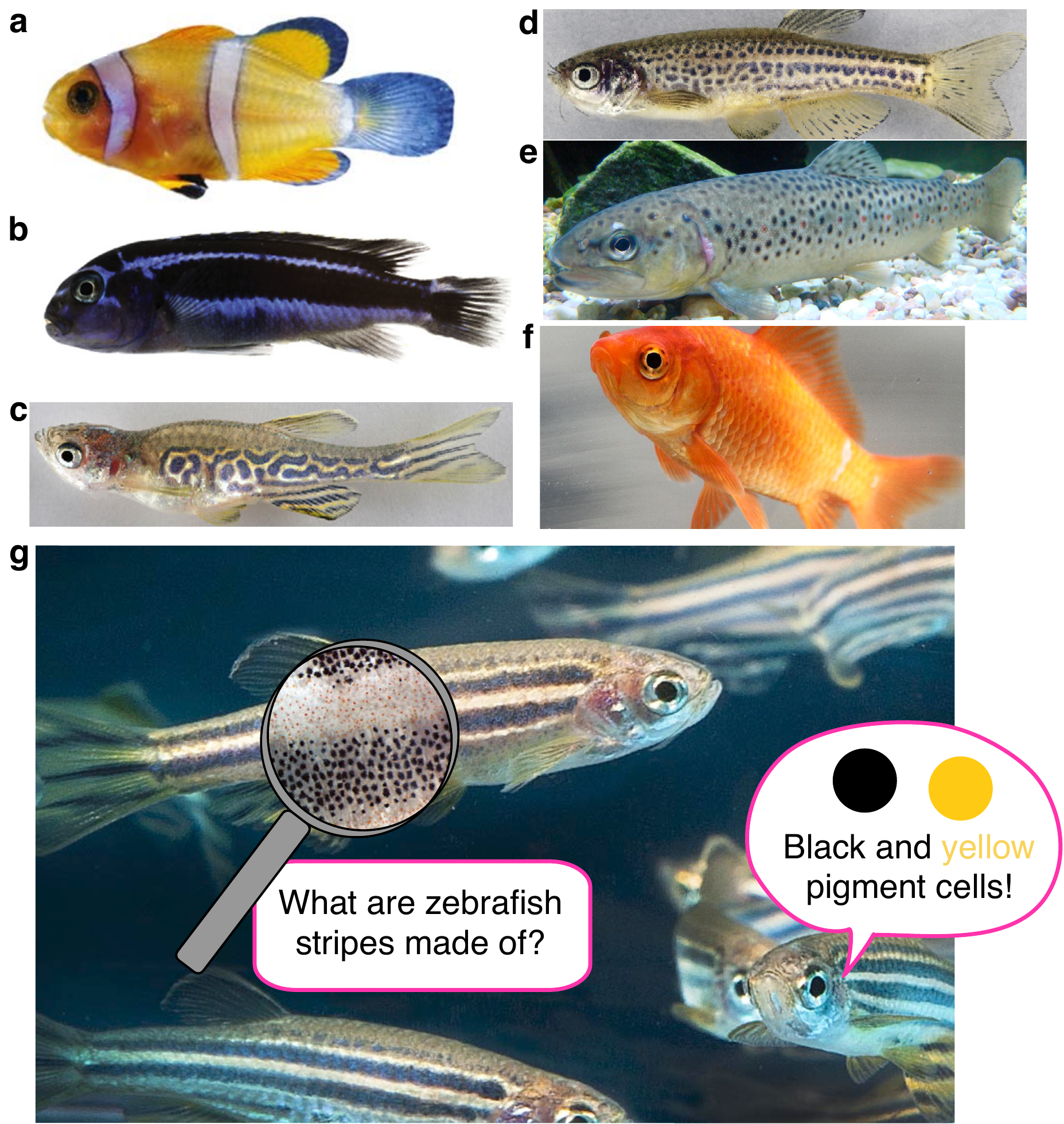}
\caption{\label{fig:zebrafishIntro} Fish have all sorts of different skin patterns, and here are some examples. Their patterns are made of \gls{pigment cell}{\color{RubineRed}s}. (Images from \cite{Frohnhofer,Salis2019,IrionRev2019,Jan} and Wikimedia Commons; see end of document for sources.)}
\end{figure}

\noindent Many types of animals have skin patterns. Zebras have stripes and leopards have spots, but fish display all sorts of patterns. When you look at Figure~\ref{fig:zebrafishIntro}, what patterns do you see? There are fish with stripes, spots, mazes, and plain patterns \cite{Frohnhofer}. Which pattern is your favorite? Zebrafish (Figure~\ref{fig:zebrafishIntro}g) are good for doing experiments with, so a lot of scientists like zebrafish best \cite{IrionRev2019,Jan}. As you might guess from the word ``zebra" in their name, zebrafish usually have stripes. \\ 

\noindent When we look at zebrafish from far away, we see black and yellow stripes. If we put a zebrafish under a microscope, however, we can see that its patterns are made up of tiny dots, or \gls{pigment cell}{\color{RubineRed}s}. Pigment cells come in different sizes and colors. Some cells are big and black (biologists call them ``melanophores"). Other cells are small and yellow (their fancy name is ``xanthophores"). There are more colors of pigment cells out there, and even humans have \gls{pigment cell}{\color{RubineRed}s} in their skin.

\section*{\color{ForestGreen}Cells follow rules to create stripes}

\noindent Okay, so fish patterns contain \gls{pigment cell}{\color{RubineRed}s}. But what's really cool is that these cells actually have the job of making the stripes \cite{Yamaguchi}! In this 
\href{https://www.pnas.org/content/104/12/4790/tab-figures-data}{\color{RubineRed}movie}\footnote{Watch a fish pattern form here: \url{https://www.pnas.org/content/104/12/4790/tab-figures-data}. Once you click the website, scroll down to ``Supporting Information" and then click ``SI Movie 1."} \cite{Yamaguchi} of a fish growing, do you see the black cells moving around? There are small yellow cells there too, but they are harder to see. You can think of each \gls{pigment cell}{\color{RubineRed}} in zebrafish skin like it's a person moving in a group. You can move around in the room, and cells can move around in the skin, too. \\

\noindent Black and yellow cells move and act in specific ways (or follow rules) to create patterns. Yellow cells following black cells is one of these rules. It's almost like they are playing \emph{tag, you're it}! You can see a black cell leading the way and a yellow cell following it in this \href{https://www.youtube.com/watch?v=0wECUnwgN8A}{\color{RubineRed}movie}\footnote{Watch yellow cells following black cells here: \url{https://www.youtube.com/watch?v=0wECUnwgN8A}} \cite{Yamanaka2014}. Crazy! 

\section*{\color{ForestGreen}How do scientists study fish patterns?}

\begin{figure}[t!]
\centering
\includegraphics[width=0.8\textwidth]{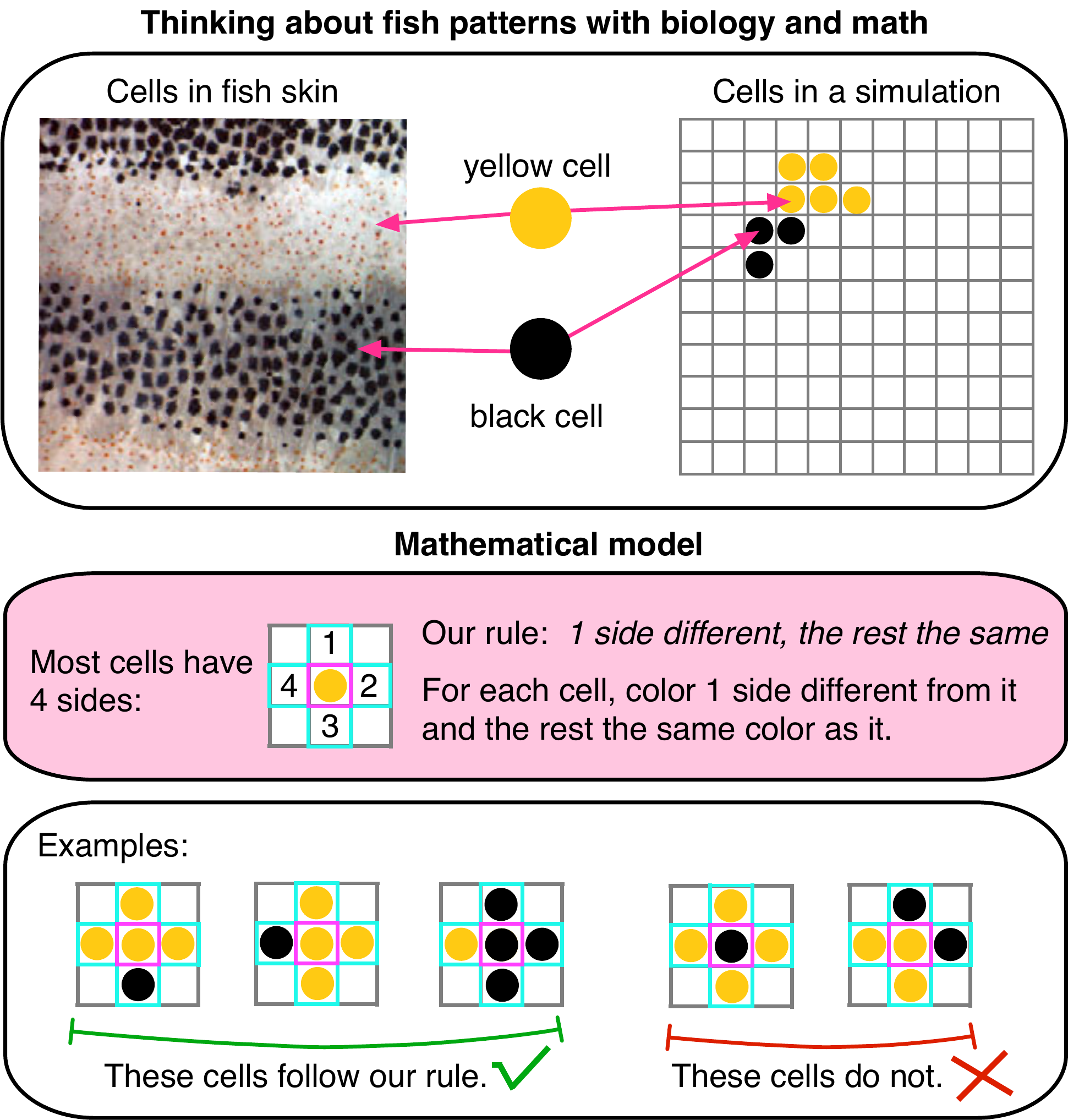}
\caption{\label{fig:zebrafishModel} Cells follow rules to create patterns. To \gls{simulate} patterns with a \gls{model} on a computer, think of fish skin as a checkerboard \cite{Bullara}. Then pick a rule for what color of cell goes in each square. Our rule is ``One side different, the rest the same." If you put your finger on a cell in the center (indicated by the pink square), one of its neighbors (indicated by green squares) should have a different color than it, and the rest should be the same color as it.}
\end{figure}

\noindent While some of the rules for how cells behave in zebrafish are already known (like yellow cells following black cells \cite{Yamanaka2014}), many rules are a mystery. Scientists are working hard to discover the rules now. Cells are tiny, and it's tough to see them and catch them interacting with each other in the fish skin. Zebrafish also take a long time to grow (a few months) \cite{Jan}, so biologists have to be patient with their experiments.  \\

\noindent This is where math and computer coding enter the picture. Math is all about patterns, rules, and puzzles, and it can be combined with biology to help figure out how cells behave \cite{Bullara,volkening2}. Mathematicians and biologists use different tools to study the same problems. Bringing their ideas together makes the mystery of how zebrafish get their stripes easier to solve.\\

\noindent While biologists study fish by looking into microscopes and doing experiments, applied mathematicians write down rules for how cells might behave \cite{Bullara, volkening2}. These rules and equations are called a \gls{model} --- they are a way of describing something in a mathematical way. It is important to remember that models are a good guess, but they could be wrong. To test their rules, mathematicians use computer coding to \gls{simulate} fish patterns. This is like growing fish on a computer! One good thing about this is that simulating fish patterns only takes a few minutes, not a few months. (But we wouldn't recommend trying to eat a simulated fish for dinner! :-)

\renewcommand{\figurename}{Table}
\setcounter{figure}{0}  
\begin{figure}[t!]
\centering
\includegraphics[width=0.7\textwidth]{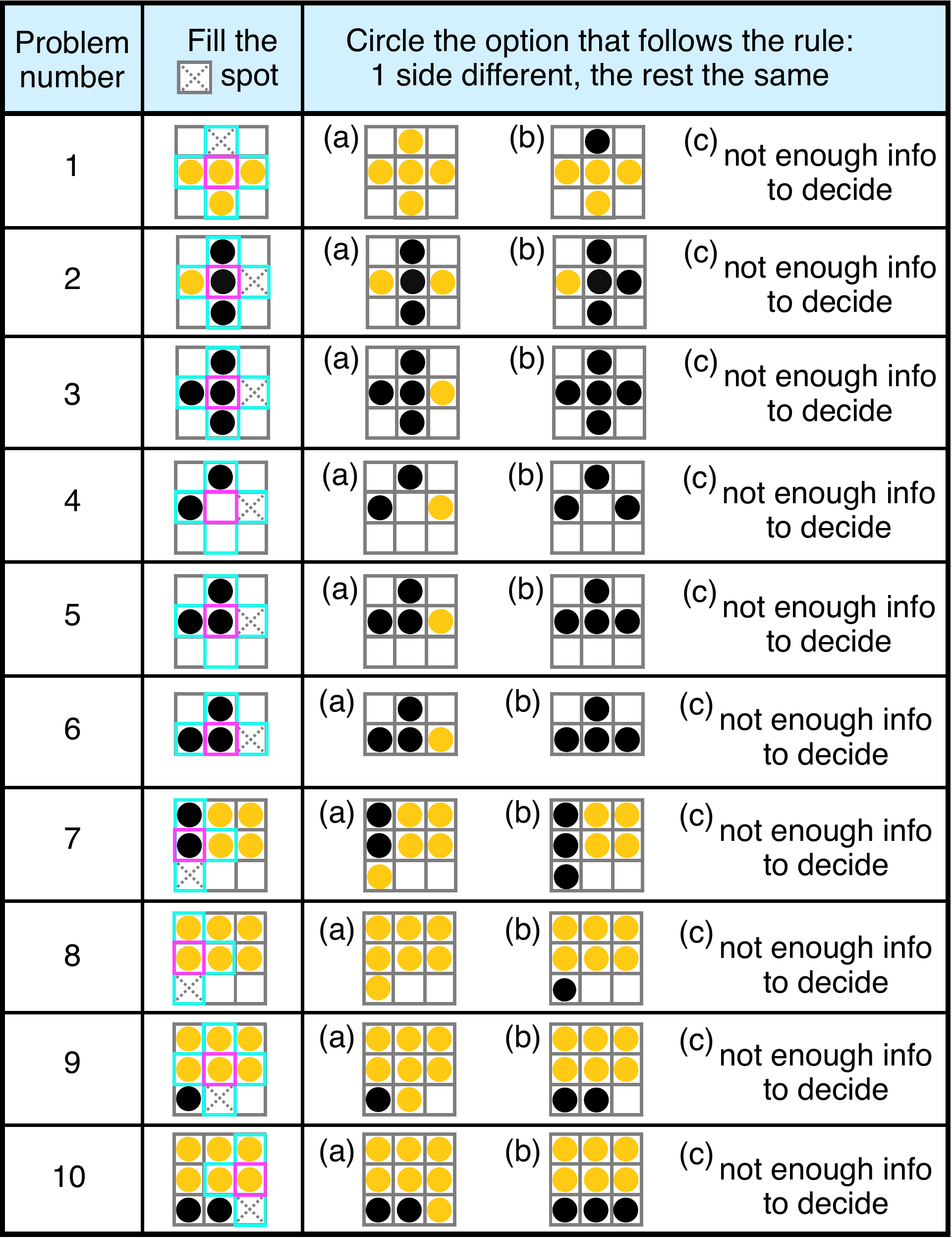}
\caption{\label{fig:zebrafishTable} Determine the option that follows the \gls{model} in Figure~\ref{fig:zebrafishModel}. Point your finger at the pink square and then use it and its green neighbors to choose the color of the X-marked square. In problem~6, the pink square only has 3 neighbors because it is at the checkerboard boundary. \textbf{Answers}: 1b, 2b, 3a, 4c, 5c (one of the focus cell's neighbors must be yellow, but it could be its right or bottom side), 6a (this cell is at the edge of the checkerboard and only has 3 sides, so we have enough information now), 7b, 8b, 9b, 10b.}
\end{figure}

\section*{\color{ForestGreen}Building a mathematical model of zebrafish stripes}

\noindent Now we are ready to build a \gls{model} of zebrafish patterns! Our first step is to describe fish skin as a checkerboard (like in Figure~\ref{fig:zebrafishModel}). Each square in our checkerboard can either be empty (white), have a yellow cell, or have a black cell.\\

\noindent To \gls{simulate} fish patterns, we need to tell the black and yellow cells where to appear in the checkerboard. In other words, we need to describe their rules of behavior. Biologists know that black cells mostly like to be near other black cells, and yellow cells like to have mostly yellow cells around them. We use this fact to build our \gls{model}. \\

\noindent In Figure~\ref{fig:zebrafishModel}, each cell in the checkerboard has 4 sides (unless the cell is at a corner or an edge). Our rule is ``One side different, the rest the same." This means that if you put your finger on a yellow cell, one of the cells that it touches should be black (different) and the other cells should be yellow (the same). You can see some examples of cells that follow this rule and cells that don't in Figure~\ref{fig:zebrafishModel}. To get some practice with this rule, try filling in the worksheet in Table~\ref{fig:zebrafishTable} (answers are at the bottom of the table).\\

\noindent Our model has only one rule, and the truth is that we are simplifying all the ways that cells actually behave in zebrafish. But that's okay! When mathematicians make models, they start with simple rules and then build more realistic ones. Simple models help scientists learn, and our model is a first step. If you'd like to see more realistic models that mathematicians are working on now, check out the scientific study \cite{volkening2}.

\section*{\color{ForestGreen}Our puzzle: Simulating zebrafish stripes}

\noindent What pattern does the rule ``One side different, the rest the same" create? You can try this out and build a fish pattern using Figure~\ref{zebrafishCraft}. If you print Figure~\ref{zebrafishCraft} and cut out the black and yellow cells, follow these steps:
\begin{itemize}
    \item \textbf{Step 1:} Zebrafish have one yellow stripe when they are young, so start by putting 2 strips of yellow cells at the top of your fish checkerboard.
    \item \textbf{Step 2:} Fill out the rest of the checkerboard by following the rule ``One side different, the rest the same." 
\end{itemize}
Start by filling out the squares at the top of the checkerboard and slowly move your finger down the fish from left to right. What pattern forms? \\

\renewcommand{\figurename}{Figure}
\setcounter{figure}{2}  
\begin{figure}[h!]
\centering
\includegraphics[width=0.82\textwidth]{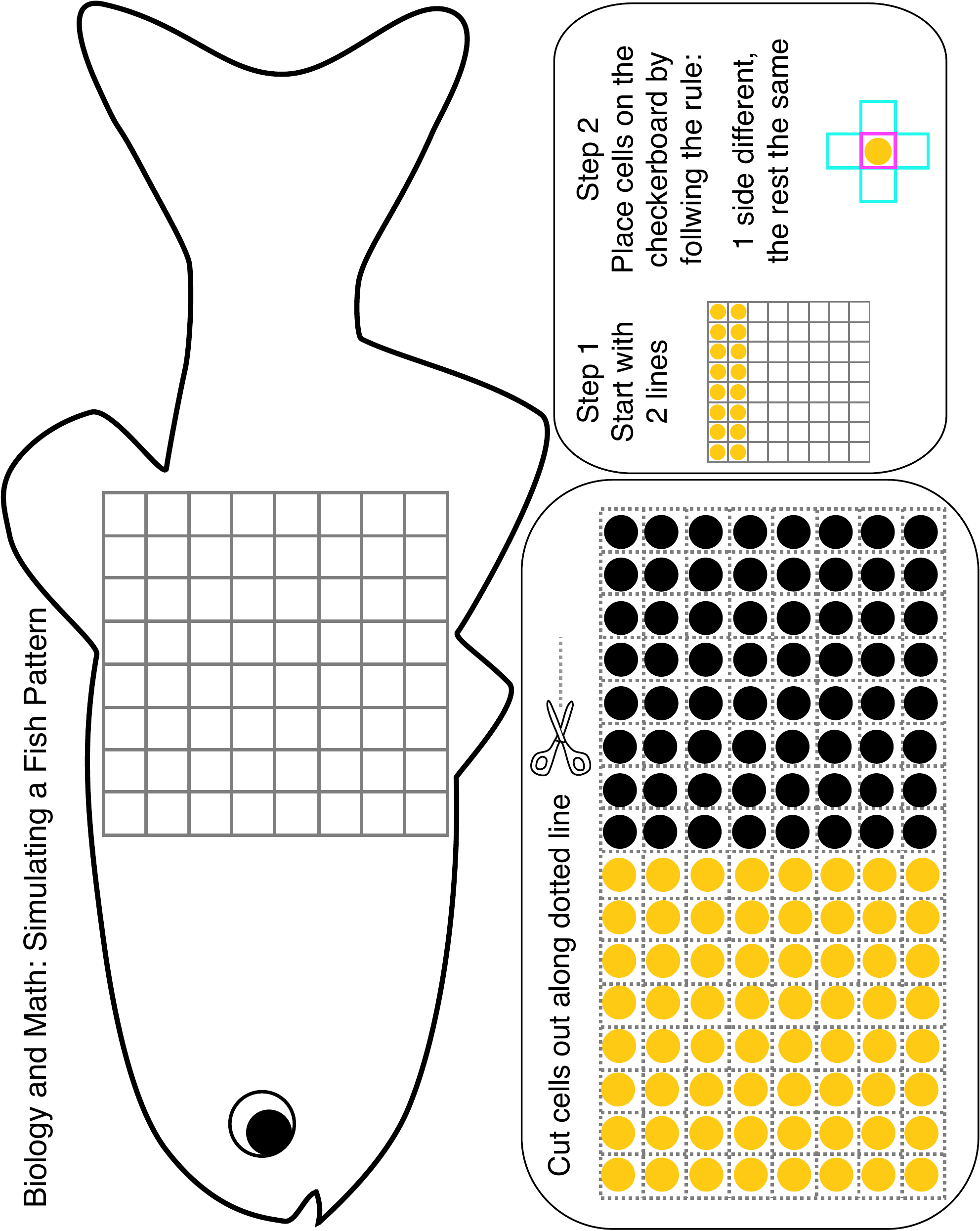}
\caption{\label{zebrafishCraft} Use this image to create a fish pattern by combining biology and math. Fill the checkerboard on the fish by following Steps 1 and 2. After you finish, print this page again. It's your turn to be the applied mathematician now --- make up your own rules to discover other patterns that you can create! How would you change Step 2 to make a goldfish? You can also see how math is combined with computer coding to \gls{simulate} different patterns at this \href{https://simulatingzebrafish.gitlab.io/zebrafish-outreach/}{\color{RubineRed}website}.}
\end{figure}

\noindent By following the steps in Figure~\ref{zebrafishCraft}, you are doing the same steps by hand that a computer would do to \gls{simulate} a fish pattern. We've used computer coding to create fish patterns --- check out our \href{https://simulatingzebrafish.gitlab.io/zebrafish-outreach/}{\color{RubineRed}website}\footnote{Simulate a fish pattern here: \url{https://simulatingzebrafish.gitlab.io/zebrafish-outreach/}} to \gls{simulate} a fish yourself!

\section*{\color{ForestGreen}Why is studying fish patterns important, anyway?}

\noindent Fish patterns might be nice to look at, but why do biologists and mathematicians study them? That's a really good question, and the answer has to do with mutant patterns, medicine, and humans.\\

\noindent Turns out that zebrafish only \emph{usually} have stripes (hence the ``zebra" part of their name). As you know, stripes form because cells follow specific rules. But what happens if cells don't follow their normal rules? This can happen if a fish has a \gls{mutation} in its \gls{genes} \cite{IrionRev2019,ParichyRev2019}. In Figure~\ref{fig:zebrafishIntro}, image (g) is not the only picture of a zebrafish. Believe it or not, images (c) and (d) are also zebrafish, but they are mutants! \\

\noindent Mutant zebrafish have different patterns because their cells follow different rules. A fish's \gls{genes} specify these rules, so mutant \gls{genes} lead to mutant rules and mutant patterns. Humans can also have \gls{mutation}{\color{RubineRed}s} in their \gls{genes}, and this can cause disease. \\

\noindent Zebrafish and humans look very different, but they actually have a lot of similar \gls{genes}. And that's why studying zebrafish is important. If we can figure out what rules cells follow to create normal and mutant patterns in fish, we might be able to better understand normal and unhealthy behavior of cells in humans down the line too.

\section*{\color{ForestGreen}Now you get to be the applied mathematician!}

\noindent Scientists are working hard to discover the ways that black and yellow cells behave in normal zebrafish, and to figure out how these cells change their behavior in mutant fish \cite{ParichyRev2019,IrionRev2019}. For mathematicians who study zebrafish, their job is to think of a fish pattern (like stripes) and then find the rules that can create that pattern \cite{Bullara,volkening2}. This is a puzzle. \\

\noindent How would you change the rule ``One side different, the rest the same" to create a plain yellow fish? What rules and starting points create diagonal stripes? Try out your own rules by printing Figure~\ref{zebrafishCraft} out again. By working on these questions, you are taking on the role of mathematician yourself! What fish patterns can you create with math?

\printglossary[nonumberlist]

\FloatBarrier

\bibliographystyle{siam}
\bibliography{reviewbib}

\paragraph{Image Sources}{Figure 1: (a) from \cite{Salis2019} with permission from John Wiley \& Sons, Copyright (2019) John Wiley \& Sons A/S; (b) from \cite{IrionRev2019} (licensed under \href{https://creativecommons.org/licenses/by-nc-nd/4.0/}{CC BY-NC-ND 4.0}); (c) from \cite{Frohnhofer} (\href{http://creativecommons.org/licenses/by/3.0}{CC-BY 3.0}); (d) from \cite{Jan} with permission from Elsevier, Copyright (2015) Elsevier Ltd.; 
(e) from Zouavman Le Zouave (\href{https://creativecommons.org/licenses/by-sa/3.0}{CC BY-SA}) via Wikimedia Commons; (f) from Bjwebb at English Wikipedia, Public domain, via Wikimedia Commons; 
(g) adapted and combined from \cite{Frohnhofer} (\href{http://creativecommons.org/licenses/by/3.0}{CC-BY 3.0}) and Oregon State University (\href{https://creativecommons.org/licenses/by-sa/2.0}{CC BY-SA}) via Wikimedia Commons. Figure 2: ``Cells in fish skin" image adapted from \cite{Frohnhofer} (\href{http://creativecommons.org/licenses/by/3.0}{CC-BY 3.0}).}

\paragraph{Conflict of Interest Statement}{ We declare no conflict of interest.}

\paragraph{Acknowledgments}{ We are grateful to Jithin George, Katelyn Joy Leisman, Niall Mangan, and Sasha Shirman for helpful feedback on this manuscript, and BS thanks Sasha for providing suggestions on JavaScript. We also thank Domenico Bullara and Yannick De Decker for publishing their JavaScript code as part of \cite{Bullara}, since it helped inspire the code that we developed for our \href{https://simulatingzebrafish.gitlab.io/zebrafish-outreach/}{website}. AV has been supported in part by the National Science Foundation under grant no.\ DMS-1764421 and by the Simons Foundation/SFARI under grant no.\ 597491-RWC.} \\ \\

\newpage

\noindent \textbf{\large{\color{ForestGreen}Author Biographies}} \\

\begin{wrapfigure}{R}{0.21\textwidth}\centering 
\vspace{-1\baselineskip}
\includegraphics[width=0.2\textwidth]{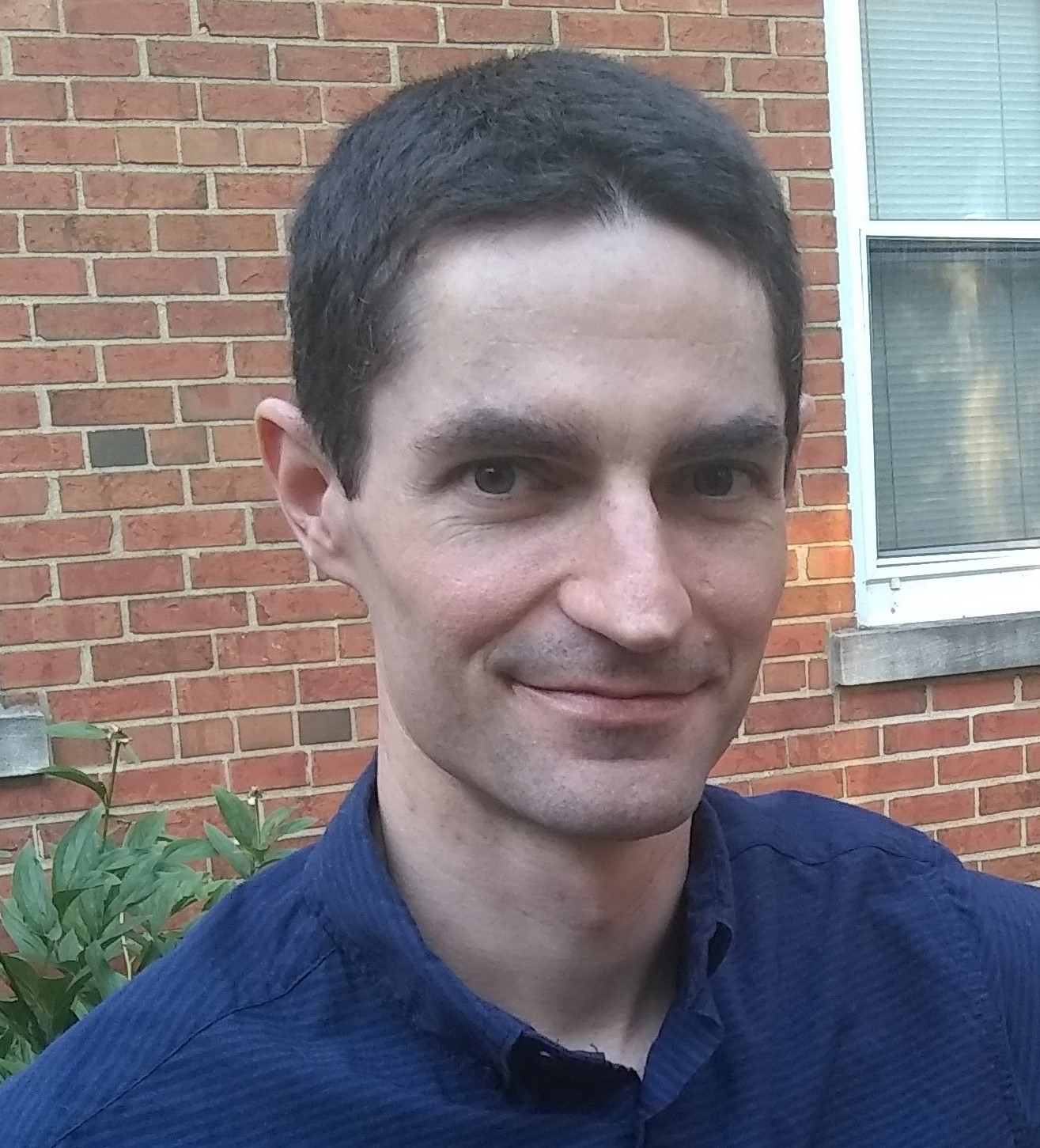}\vspace{-2\baselineskip}
\end{wrapfigure}
\noindent \textbf{\color{RubineRed}Blake Shirman}{ is a Master's student at DePaul University, and he hopes to one day get a Ph.D. in pure math. He spends his time looking at logic puzzles and games to see if math can make things easier, sometimes in fun and unexpected ways! One of his favorite puzzles is trying to predict cell patterns (like how to place checkers on a checkerboard with special rules about where they go). It wasn't until he met Alexandria that Blake learned diagrams like this were being used to describe the vibrant stripes and spots on many of our undersea friends!}

\vspace{3\baselineskip}

\begin{wrapfigure}{L}{0.21\textwidth}\centering \vspace{-1\baselineskip}
\includegraphics[width=0.2\textwidth]{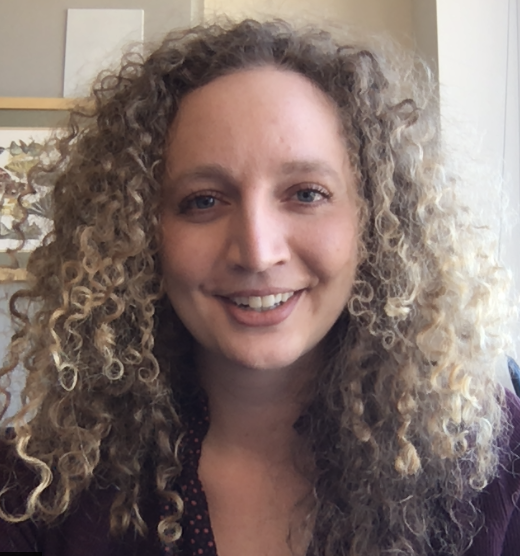}
\end{wrapfigure}
\noindent \textbf{\color{RubineRed}Alexandria Volkening}{ is an Assistant Professor in the Department of Mathematics at Purdue University, and she got her Ph.D. in applied mathematics from Brown University. She studies all sorts of stuff with math (zebrafish patterns, of course, but also political elections, social media, and crowds of people --- turns out lots of things follow rules of behavior that can be described mathematically!). Her favorite fish to simulate (so far) is zebrafish, but her favorite fish for dinner is salmon. And no, she doesn't have any pet zebrafish, so she always enjoys visiting biologists' labs to see them!}

\end{document}